\documentstyle[12pt]{article}

\def \cp89{{\it CP Violation,} edited by C. Jarlskog (World Scientific,
Singapore, 1989)}

\def \f79{{\it Proceedings of the 1979 International Symposium on Lepton and
Photon Interactions at High Energies,} Fermilab, August 23-29, 1979, ed. by
T. B. W. Kirk and H. D. I. Abarbanel (Fermi National Accelerator Laboratory,
Batavia, IL, 1979}
\def \hb87{{\it Proceeding of the 1987 International Symposium on Lepton and
Photon Interactions at High Energies,} Hamburg, 1987, ed. by W. Bartel
and R. R\"uckl (Nucl. Phys. B, Proc. Suppl., vol. 3) (North-Holland,
Amsterdam, 1988)}

\def \ichep72{{\it Proceedings of the XVI International Conference on High
Energy Physics}, Chicago and Batavia, Illinois, Sept. 6 -- 13, 1972,
edited by J. D. Jackson, A. Roberts, and R. Donaldson (Fermilab, Batavia,
IL, 1972)}

\def \ite{{\it et al.}}
\def \lkl87{{\it Selected Topics in Electroweak Interactions} (Proceedings of
the Second Lake Louise Institute on New Frontiers in Particle Physics, 15 --
21 February, 1987), edited by J. M. Cameron \ite~(World Scientific, Singapore,
1987)}
\def \ky85{{\it Proceedings of the International Symposium on Lepton and
Photon Interactions at High Energy,} Kyoto, Aug.~19-24, 1985, edited by M.
Konuma and K. Takahashi (Kyoto Univ., Kyoto, 1985)}

\def \si90{25th International Conference on High Energy Physics, Singapore,
Aug. 2-8, 1990}
\def \slc87{{\it Proceedings of the Salt Lake City Meeting} (Division of
Particles and Fields, American Physical Society, Salt Lake City, Utah, 1987),
ed. by C. DeTar and J. S. Ball (World Scientific, Singapore, 1987)}
\def \slac89{{\it Proceedings of the XIVth International Symposium on
Lepton and Photon Interactions,} Stanford, California, 1989, edited by M.
Riordan (World Scientific, Singapore, 1990)}
\def \smass82{{\it Proceedings of the 1982 DPF Summer Study on Elementary
Particle Physics and Future Facilities}, Snowmass, Colorado, edited by R.
Donaldson, R. Gustafson, and F. Paige (World Scientific, Singapore, 1982)}
\def \smass90{{\it Research Directions for the Decade} (Proceedings of the
1990 Summer Study on High Energy Physics, June 25--July 13, Snowmass,
Colorado),
edited by E. L. Berger (World Scientific, Singapore, 1992)}
\def \tasi90{{\it Testing the Standard Model} (Proceedings of the 1990
Theoretical Advanced Study Institute in Elementary Particle Physics, Boulder,
Colorado, 3--27 June, 1990), edited by M. Cveti\v{c} and P. Langacker
(World Scientific, Singapore, 1991)}

\newcommand{\pdrv}[2] { {\partial{#1}\over{\partial{#2}}} }
\newcommand{\pdrvsl}[2]{ \partial{#1}/{\partial{#2}} }
\newcommand{\pdrvbl}[1]{ {\partial\over{\partial{#1}}} }
\newcommand{\vdrv}[2] { {{\delta #1}\over{\delta #2 }} }
\newcommand{\vdrvsl}[2] {{\delta #1 }/{\delta #2 }}
\newcommand{\deriv}[2] { {d{#1}\over{d{#2}}} }
\newcommand{\integral}[1] { {\int_{-\infty}^{\infty} {#1} \, dx} }
\begin{document}
\begin{titlepage}
{\large
	\hspace*{\fill} EFI 93-20\\
	\hspace*{\fill} hep-th/9304139\\
	\hspace*{\fill} April 1993\\
}
\bigskip
\begin{center}
	{\Large \bf Supersymmetric quantum mechanics and the
	Korteweg-de Vries hierarchy}\\
\end{center}
\bigskip
\begin{center}
	{\large Aaron K. Grant and Jonathan L. Rosner\\}
	Enrico Fermi Institute and Department of Physics\\
	University of Chicago, Chicago, IL 60637\\
\end{center}
\medskip
\centerline{April 1993}
\bigskip
\begin{abstract}
The connection between supersymmetric quantum mechanics and the Korteweg-
de Vries (KdV) equation is discussed, with particular emphasis on the
KdV conservation laws.  It is shown that supersymmetric quantum mechanics
aids in the derivation of the conservation laws, and gives some insight
into the Miura transformation that converts the KdV equation into
the modified KdV equation.  The construction of the $\tau$-function by
means of supersymmetric quantum mechanics is discussed.

\end{abstract}
\end{titlepage}
\renewcommand{\thesection}{\Roman{section}}
\renewcommand{\thetable}{\Roman{table}}

\section{INTRODUCTION}

It is well known that both the Korteweg-de Vries (KdV) equation [1] and
supersymmetric quantum mechanics [2] have intimate connections to the inverse
scattering problem [3-6].  In Ref.~[4] it was shown that supersymmetric
quantum mechanics can be used to construct reflectionless one-dimensional
potentials with arbitrarily prescribed bound state energies.  On the other
hand, one can solve the KdV equation by means of the inverse scattering
transform [1], wherein one associates a solution of KdV with a Schr\"odinger
potential. Further links between supersymmetric quantum mechanics and the KdV
equation have been found in connection with B\"acklund transformations [4],
which can be used either to add a soliton to a solution of KdV, or to add a
bound state to a one-dimensional potential.  It has been shown that
supersymmetric quantum mechanics allows one to obtain an expression for the
$\tau$- function that comes up in solving the KdV equation [5], and to obtain
one- and two-soliton solutions for an infinite family of equations related to
KdV [6]. In addition, it has been noted [7] that the change of variable
connecting the modified KdV equation with the usual KdV equation comes up quite
naturally in factorizing the Schr\"odinger equation.

We have two parallel ways of adding a soliton to an already existing
$n$-soliton solution.  (1)  We can perform a transformation within
supersymmetric quantum mechanics [3-6].  (2) We can construct a $\tau$-function
for the $n$-soliton solution which, when acted upon by a suitable vertex
operator, yields a $\tau$-function for the $n+1$-soliton solution [8].  Our
purpose in the present paper is to explore the relation between these
two methods.

In Sec. II we review the construction of one-dimensional potentials using
supersymmetric quantum mechanics.  In Sec. III we derive conservation laws for
deformations of a potential that leave the bound state energies unchanged,
while Sec. IV introduces a hierarchy of equations of the KdV type as a means of
generating such deformations. In Sec. V, we introduce the $\tau$-function and
the vertex operator, and present a method for constructing the $n$-soliton
$\tau$-function using supersymmetric quantum mechanics. We conclude in Sec.~VI.

\section{SUPERSYMMETRIC QUANTUM MECHANICS}
In this section we consider the construction of reflectionless one-dimensional
potentials by means of supersymmetric quantum mechanics. We begin by writing
the one-dimensional Schr\"odinger Hamiltonian
\begin{equation}
H_{+}=-{d^2\over{dx^2}}+V_{+}(x)
\end{equation}
in factorized form as
\begin{equation}
\label{factorization}
H_{+}=A^{\dagger}A,
\end{equation}
where A is given by
\begin{equation}
A=-{d\over{dx}}+f(x).
\end{equation}
In order that Eq.~(\ref{factorization}) hold, $f(x)$ must obey the
Ricatti equation
\begin{equation}
f^2+f'=V_{+}.
\end{equation}
The form of Eq.~(\ref{factorization}) suggests the introduction of a `partner'
Hamiltonian $H_{-}$ given by
\begin{equation}
H_{-}=-{d^2\over{dx^2}}+V_{-}(x)=AA^{\dagger}.
\end{equation}
The potentials $V_{\pm}$ are given in terms of $f$ by $V_{\pm}=f^2 \pm f'$.

It turns out that there are certain interesting relationships between the
spectra of $H_{+}$ and $H_{-}$.  For suppose that $\psi_{+}$ is an
eigenfunction
$H_{+}$.  We then have
\begin{equation}
A^{\dagger}A\psi_{+}=E_{+}\psi_{+}.
\end{equation}
Multiplying on the left by $A$ gives
\begin{equation}
AA^{\dagger}[A\psi_{+}]=E_{+}[A\psi_{+}].
\end{equation}
{}From this it follows either that $\psi_{-}\equiv A\psi_{+}$ is an
eigenfunction
of $H_{-}$ with eigenvalue $E_{+}$, or that $\psi_{-}\equiv 0$.  Now the
latter of these two possibilities implies that
\begin{equation}
0=\langle\psi_{-}|\psi_{-}\rangle=\langle A\psi_{+}|A\psi_{+}\rangle
	=\langle\psi_{+}|A^{\dagger}A\psi_{+}\rangle=E_{+}\langle\psi_{+}|
	\psi_{+}\rangle,
\end{equation}
so that $\psi_{-}\equiv 0$ only when $E_{+}=0$.  It follows that
$H_{+}$ and $H_{-}$ have the same spectra,
apart from the single $E=0$
eigenvalue which is present in the spectrum of $H_{+}$, but absent from the
spectrum of $H_{-}$.  A similar argument shows that if $\psi_{-}$ is an
eigenfunction of $H_{-}$, then $\psi_{+}=A^{\dagger}\psi_{-}$ is an
eigenfunction of $H_{+}$.

The factorization method can be used to construct reflectionless potentials
possessing an arbitrary spectrum of bound states.  To see this, suppose
we have a Hamiltonian $H^{(1)}$ with potential $V^{(1)}$ having $n$ eigenvalues
$E=E_1,~E_2,\dots,~ E_{n}$.  We may construct from $V^{(1)}$ a potential
$V^{(2)}$ having the $n$ eigenvalues $E_{k}$, plus one more eigenvalue
$E_{n+1}$. To do this, first choose the zero of energy such that
$V^{(1)}\rightarrow 0$ as $x\rightarrow\pm\infty$, and define $V_{-}$ by
$V_{-}=V^{(1)}-E_{n+1}$.  We factorize the corresponding Schr\"odinger equation
as above:  define $f(x)$ by
\begin{equation}
f^2-f'=V_{-},
\end{equation}
and construct the partner potential
\begin{equation}
V_{+}=f^2+f'.
\end{equation}
The potential $V_{-}$ has the $n$ eigenvalues
$E_{1}-E_{n+1},~E_{2}-E_{n+1},\dots,~E_{n}-E_{n+1}$, while $V_{+}$ has $n+1$
eigenvalues consisting of the $n$ eigenvalues of $V_{-}$, plus the additional
eigenvalue $E=0$. Defining $V^{(2)}=V_{+}+E_{n+1}$ yields the desired
potential, possessing the $n+1$ eigenvalues $E_1,~E_2,\dots,~E_{n+1}$.  In this
way, one can start from the constant potential $V^{(1)}=0$ and by iteration
build up a potential possessing an arbitrary spectrum of eigenvalues.

As an example of this procedure we construct a potential possessing a single
bound state with energy $E=-\kappa^2$.  We begin from $V^{(1)}=0$, so that
$V_{-}=\kappa^2$, and $f$ obeys
\begin{equation}
f^2-f'=\kappa^2.
\end{equation}
This may be linearized by the substitution $f=-w'/w$, yielding
\begin{equation}
\label{f}
f=-\kappa\tanh\kappa(x-x_{0}),
\end{equation}
so that $V_{+}$ is given by
\begin{equation}
V_{+}=\kappa^2\biggl{[} 1-2{\rm sech}^2 \kappa(x-x_{0}) \biggr{]},
\end{equation}
and the desired potential, possessing a single bound state, is given by
\begin{equation}
V^{(2)}=-2\kappa^2{\rm sech}^2\kappa(x-x_{0}).
\end{equation}

To see that $V^{(2)}$ is reflectionless, consider a plane wave solution
for the potential $V_{-}$, $\psi_{-}(x)=e^{ikx}$.  The corresponding
solution for the potential $V_{+}$ is given by $\psi_{+}=A^{\dagger}\psi_{-}$.
{}From
Eq.~(\ref{f}),  it follows that $\psi_{+}$
behaves asymptotically like $(ik\mp\kappa)e^{ikx}$ as $x\rightarrow\pm\infty$.
Since no term proportional to $e^{-ikx}$ arises, the potential is
reflectionless. Defining the reflection and transmission coefficients
$R$ and $T$ by
\begin{equation}
\label{deftransmission}
\psi(x)\rightarrow\cases{
			e^{ikx}+R(k)e^{-ikx},  & $x \rightarrow -\infty$;\cr
\cr
			T(k)e^{ikx}, & $x \rightarrow +\infty$,\cr}
\end{equation}
we find that $R=0$, while
\begin{equation}
T(k)={{ik-\kappa}\over{ik+\kappa}}.
\end{equation}
By repeating the above procedure, it is possible
to construct potentials possessing
arbitrarily many bound states.  It is shown in the Appendix that such
potentials are reflectionless, and that furthermore they have transmission
coefficients given by
\begin{equation}
\label{transmission}
T(k)=\prod_{i=1}^{N} {{ik-\kappa_i}\over{ik+\kappa_i}},
\end{equation}
where the $N$ bound states have energies $E_i=-\kappa_i^2$.

\section{CONSERVATION LAWS FOR ISOSPECTRAL DEFORMATIONS OF A POTENTIAL}

Consider again the one-dimensional Schr\"odinger equation, with a
reflectionless potential $V(x)=-u(x)$:
\begin{equation}
\label{Seqn}
\psi_{xx}+[k^2+u(x)]\psi=0.
\end{equation}
We wish to derive certain quantities which will be conserved under any
deformation of the potential which leaves the spectrum of bound states
unchanged.  These conserved quantites may be obtained via an asymptotic
expansion of the transmission coefficient for large values of $k$.  We
write the wavefunction as
\begin{equation}
\label{psi}
\psi(x)=\exp \biggl{[}ikx+\int_{-\infty}^{x} \phi(x') \, dx'\biggr{]}
\end{equation}
and expand $\phi$ as
\begin{equation}
\phi(x)=\sum_{n=1}^{\infty} {f_{n}(x)\over{(2ik)^n}}.
\end{equation}
Substitution of (\ref{psi}) into (\ref{Seqn}) yields the recursion relation
\begin{equation}
f_{n+1}=-f_{n,x} - \sum_{k=1}^{n-1} f_k f_{n-k},
\end{equation}
for $n \ge 1$, while $f_1=-u(x)$.  The first few terms in the series are
$$
\displaylines{
f_1=-u,~f_2=u_x,~f_3=-u_{xx}-u^2,~f_4=(u_{xx}+2u^2)_x,~{\rm and} \cr
f_5=-(u_{xx}+3u^2)_{xx}+u_x^2-2u^3.
               }
$$
Taking the limit $x\rightarrow\pm\infty$ in Eq.~(\ref{psi}) and comparing with
Eq.~(\ref{deftransmission}) shows that the transmission coefficient
given in Eq.~(\ref{transmission}) may be written in two ways:
\begin{equation}
\label{Tseries}
\log T(k)=
	\sum_{n=1}^{\infty} {1\over{(2ik)^n}}\int_{-\infty}^{\infty} f_n(x)\,dx
	=\sum_{i=1}^{N}\biggl{[}\log(ik-\kappa_i)-\log(ik+\kappa_i)\biggr{]}.
\end{equation}
Expanding the quantity on the far right in a power series in $1/k$ and
equating coefficients of $1/k^n$ gives
\begin{equation}
\label{proof}
\int_{-\infty}^{\infty} f_{n}(x)\,dx=\cases{
		0 & if $n$ is even;\cr
				   \cr
		- {2^{n+1}\over{n}} \sum_{i=1}^{N} \kappa_{i}^n
		  & if $n$ is odd.\cr}
\end{equation}
The integrals vanish for even $n$ because the integrands in this case are total
derivatives of a function having the same values at $x = \pm \infty$, as we see
for $n=2$ and $n=4$ in the list above.  It follows from Eq.~(\ref{proof}) that
the integrals of the functions $f_n$ are constants of the motion for any
spectrum-preserving deformation of the potential $u(x)$: since the eigenvalues
$\kappa_i^2$ are constant for such a deformation, the integrals must also be
constant.  These integrals are precisely the KdV Hamiltonians, up to an overall
multiplicative factor.  The choice of this multiplicative factor is simply a
matter of convenience, so we choose to define the Hamiltonians as
\begin{equation}
H_{2n+1}[u]=-{1\over{2^{2n+1}}}\int_{-\infty}^{\infty} f_{2n+3}(x)\, dx.
\end{equation}

A specific form for $f_n(x)$ may be written for the single-soliton solution by
noting that for a soliton with $x_0 = 0$,
$$
\displaylines{\psi = A^{\dagger} e^{ikx} \sim \left( {ik - \kappa \tanh \kappa
x
\over ik + \kappa} \right) e^{ikx} \cr
= \exp(ikx + \log[ik - \kappa \tanh \kappa x] -\log[ik + \kappa])
\equiv \exp[ikx + \int_\infty^x \phi(x') dx']~~~.}
$$
Expanding
\begin{equation}
\phi(x) = \sum_{n=1}^{\infty} {f_{n}(x)\over{(2ik)^n}} =
(d/dx) (\log[ik - \kappa \tanh \kappa x] -\log[ik + \kappa])
\end{equation}
gives the result
\begin{equation}
f_n = - {(2 \kappa)^n \over n} {d \over dx} \tanh^n \kappa x~~~.
\end{equation}

\section{EVOLUTION EQUATIONS}

We can use the results of the previous two sections to derive a set of spectrum
preserving evolution equations for the potential $u(x)$. We know from the last
section that any such evolution equation must conserve the transmission
coefficient $T(k)$, and, consequently, must also conserve each of the
Hamiltonians $H_{2n+1}$.  A natural choice for an evolution equation conserving
a specific Hamiltonian $H_{2k+1}$ is a member of the Korteweg-de Vries
hierarchy, which may be written in the form
\begin{equation}
\label{kdv}
u_t=\pdrvbl{x} \vdrv{H_{2k+1}}{u}.
\end{equation}
We have introduced the variational derivative $\vdrvsl{}{u}$, defined
by
\begin{equation}
\delta H[u]=\int_{-\infty}^{\infty} \vdrv{H}{u} \delta u \,dx,
\end{equation}
so that if $H[u]=\int_{-\infty}^{\infty} h(u,u_x,\dots)\,dx$,
\begin{equation}
\vdrv{H}{u}=\sum_{k=0}^{\infty} \biggl{(}-{d\over{dx}}\biggr{)}^k
		\pdrv{h}{u^{(k)}},
\end{equation}
where $u^{(k)}$ is the $k^{th}$ derivative of $u(x,t)$ with respect to $x$.
Eq.~(\ref{kdv}) automatically conserves the Hamiltonian $H_{2k+1}$ since
\begin{equation}
\deriv{H_{2k+1}}{t}=\integral{ \vdrv{H_{2k+1}}{u} u_{t} }
		   =\integral{ {1\over{2}}\deriv{}{x}\biggl{(}
		    \vdrv{H_{2k+1}}{u} \biggr{)}^2  }	=0.
\end{equation}
The proof that Eq.~(\ref{kdv}) also conserves {\it all} of the Hamiltonians
is somewhat more subtle.

To show that Eq.~(\ref{kdv}) conserves all of the Hamiltonians, we first
derive a new expression for the transmission coefficient $T(k)$.  We
again employ a factorization method to write the Schr\"odinger equation
in the form
\begin{equation}
\biggl{(} -{{d^2}\over{dx^2}} -[k^2+u]\biggr{)}\psi=AA^{\dagger}\psi=0,
\end{equation}
where $A=-\deriv{}{x}+f$, and
\begin{equation}
f_{x}-f^2-k^2=u.
\end{equation}
Since $AA^\dagger \psi=0$ implies $A^\dagger\psi=0$, we find that
\begin{equation}
f=-\deriv{}{x}\log\psi.
\end{equation}
Since $\psi\rightarrow T(k)\exp(ikx)$ as $x\rightarrow\infty$, we conclude
that the transmission coefficient may be written as
\begin{equation}
\label{Tintegral}
\log T(k)=-\integral{(ik+f)}.
\end{equation}
It follows that if the evolution equation for $f$ corresponding to
Eq.~(\ref{kdv}) has the form $f_t = \pdrvsl{J}{x}$, then $T(k)$, and all of
the Hamiltonians, will be conserved.  It turns out that this is
in fact the case.  To see this, we make use of the following observations:
First, we note that the variational derivatives of the Hamiltonians
obey the recursion relation [1]
\begin{equation}
\label{recursion}
\pdrv{}{x} \vdrv{H_{2k+1}}{u}={1\over{4}}\biggl{(}
	{\partial^3\over{\partial x^3}}+2u_x +4u{\partial\over{\partial x}}
	\biggr{)} \vdrv{H_{2k-1}}{u}\equiv {1\over{4}}M\vdrv{H_{2k-1}}{u}.
\end{equation}
Next, we observe that the linear operator $M$ may be written as
\begin{equation}
M=-\biggl{(}\pdrv{}{x}-2f\biggl{)}\pdrv{}{x}
	\biggl{(}-\pdrv{}{x}-2f\biggr{)}-4k^2\pdrv{}{x}.
\end{equation}
Finally, note that the variational derivatives $\vdrvsl{H}{u}$ and
$\vdrvsl{H}{f}$ are related in a simple way: we have
\begin{eqnarray}
\delta H & =& \integral{\vdrv{H}{u}\delta u} =
	 \integral{\vdrv{H}{u}\biggl{(}\pdrv{}{x}-2f\biggr{)} \delta f} =
	 \nonumber \\
	&=& \integral{\biggl{(}-\pdrv{}{x}-2f\biggr{)}\vdrv{H}{u} \delta f}=
	 \integral{\vdrv{H}{f}\delta f},
\end{eqnarray}
so that
\begin{equation}
\label{vderivreln}
\vdrv{H}{f}=\biggl{(}-\pdrv{}{x}-2f\biggr{)}\vdrv{H}{u}.
\end{equation}
We can now derive an evolution equation for $f(x,t)$.  Beginning from
Eq.~(\ref{kdv}), we find
\begin{eqnarray}
\label{monster}
u_t &=& \biggl{(}\pdrv{}{x}-2f\biggl{)} f_t = \pdrv{}{x} \vdrv{H_{2k+1}}{u}
	\nonumber\\
    &=&-\frac{1}{4}\biggl{(}\pdrv{}{x}-2f\biggl{)}\pdrv{}{x}
        \biggl{(}-\pdrv{}{x}-2f\biggr{)}\vdrv{H_{2k-1}}{u}-k^2\pdrv{}{x}
	\vdrv{H_{2k-1}}{u}
	\nonumber\\
    &=&\biggl{(}\pdrv{}{x}-2f\biggr{)}\pdrv{}{x}
	\biggl{(}-\frac{1}{4}\vdrv{H_{2k-1}}{f}
	+\frac{k^2}{4}\vdrv{H_{2k-3}}{f}\biggr{)}+k^4\pdrv{}{x}
        \vdrv{H_{2k-3}}{u}.
\end{eqnarray}
We have made repeated use of Eq.~(\ref{recursion}) to lower the index $k$, and
Eq.~(\ref{vderivreln}) to express $\vdrvsl{}{u}$ in terms of $\vdrvsl{}{f}$.
We
may continue lowering $k$ in this manner until the ``remainder'' term (i.e.
the term which is not multiplied by $(\pdrvsl{}{x}-2f)$ on the right hand side
of Eq.~(\ref{monster})) has the form $\pdrvsl{}{x}(\vdrvsl{H_{-1}}{u})$.  At
this point, the remainder vanishes and the sequence terminates.  Hence the
evolution equation for $f$ has the form
\begin{equation}
\label{ev1}
 \biggl{(}\pdrv{}{x}-2f\biggl{)} \biggl{(} f_t - \frac{1}{4}\pdrv{}{x}
	\sum_{l=0}^{k} (-1)^{l+1} k^{2l}\vdrv{H_{2(k-l)-1}}{f}  \biggr{)}=0.
\end{equation}
An argument due to Miura, Kruskal and Gardner [9] can now be used to show that
Eq.~(\ref{ev1}) implies
\begin{equation}
\label{ev2}
 f_t = \pdrv{}{x} \biggl{(}\frac{1}{4}
        \sum_{l=0}^{k} (-1)^{l+1}k^{2l}\vdrv{H_{2(k-l)-1}}{f}\biggr{)}.
\end{equation}
The argument runs as follows:  We know from Sec. III that $f$, which is
the logarithmic derivative of the wavefunction, may be expanded in an
asymptotic series of the form
\begin{equation}
f=ik+\sum_{n=1}^{\infty}\frac{a_n (x,t)}{(2ik)^n}.
\end{equation}
Consequently, the quantity
\begin{equation}
F\equiv f_t - \frac{1}{4}\pdrv{}{x}
        \sum_{l=0}^{k} (-1)^{l+1}k^{2l}\vdrv{H_{2(k-l)-1}}{f}
\end{equation}
may also be written as a series in the form
\begin{equation}
F=\sum_{n=-\infty}^{N} b_n (x,t) (2ik)^n.
\end{equation}
Substituting these expansions into Eq.~(\ref{ev1}) and equating the
coefficients of each power of $k$ to zero, we find that the coefficient
of $k^{N+1}$ is simply $b_N$, so that $b_N=0$.  But this in turn
implies that $b_{N-1}=b_{N-2}=b_{N-3}=\dots=0$.  As a result, we conclude
that $F\equiv 0$, and Eq.~(\ref{ev2}) follows.

The evolution equation (\ref{ev2}) is of the form
\begin{equation}
f_t= \pdrv{}{x} J(x).
\end{equation}
{}From Eq.~(\ref{Tintegral}) it then follows that the transmission coefficient
is a constant of the motion.  Since the coefficients of the asymptotic
expansion (\ref{Tseries}) of $\log T(k)$ are just constant multiples
of the KdV Hamiltonians,
it follows that each evolution equation of the form (\ref{kdv})
conserves all of the Hamiltonians.
Consequently, each such equation also
preserves the eigenvalue spectrum of the potential $u(x)$.

The above formalism permits a simple derivation of the time dependence
of the one soliton solution of KdV under the
evolution equations (\ref{kdv}).  The fact that each of these evolution
equations conserves {\it all} of the Hamiltonians implies that
the various evolutions commute with one another [1].  As a result,
the function $u$ may be thought of as depending on an infinite
number of time variables, with the dependence on the various times
fixed by
\begin{equation}
\label{manytimes}
u_{t_{2k+1}}=\pdrv{}{x} \vdrv{H_{2k+1}}{u}.
\end{equation}
We illustrate this by considering the one soliton solution of the KdV
hierarchy. We make the ansatz $u(x,t_1,t_3,\dots)=2\kappa^2
{\rm sech}^2(\kappa x +
\sum_{0}^{\infty} \alpha_{2l+1} t_{2l+1})$, and subtitute into
the evolution equations (\ref{manytimes}) to determine the constants
$\alpha_i$.
We find that Eq.~(\ref{manytimes}) with $k=0$ gives, after some manipulation,
\begin{equation}
2\kappa\alpha_{1}\pdrv{}{x}
{\rm sech}^2(\kappa x +\sum_{l=0}^{\infty} \alpha_{2l+1} t_{2l+1})=
2\kappa^{2}\pdrv{}{x}
{\rm sech}^2(\kappa x +\sum_{l=0}^{\infty} \alpha_{2l+1} t_{2l+1}),
\end{equation}
so that $\alpha_1=\kappa$.  Using the recursion relation (\ref{recursion}),
it can be shown that
\begin{equation}
\pdrv{}{x} \vdrv{H_{2k+1}}{u} = \kappa^{2} \pdrv{}{x} \vdrv{H_{2k-1}}{u}.
\end{equation}
{}From this all of the constants $\alpha_{i}$ can be determined from
$\alpha_1$.
The result is $\alpha_{2l+1}=\kappa^{2l+1}$, so that the full
time dependence of the one-soliton solution is given by
\begin{equation}
\label{onesoliton}
u(x,t_1,t_3,\dots)=2\kappa^2 {\rm sech}^2 (\kappa x +
\kappa t_1 + \kappa^3 t_3 + \kappa^5 t_5 + \dots).
\end{equation}
An equivalent result has been given in [6].

\section{THE $\tau$-FUNCTION AND THE VERTEX OPERATOR}

In this section we discuss the construction of reflectionless potentials and
multisoliton solutions of KdV using the $\tau$-function.  A generic
potential can be expressed in terms of the $\tau$-function by
\begin{equation}
V(x)=-2 {d^{2}\over{dx^2}}\log \tau (x).
\end{equation}
The $\tau$-function can be constructed in either of two ways:  First, we can
use supersymmetric quantum mechanics to build up the $\tau$-function from
a sort of ``vacuum state'' using the techniques discussed above.
Alternatively,
we can use the vertex operator [1,8] to construct the desired $\tau$-function.

To build up the $\tau$-function using supersymmetric quantum mechanics, we
consider a sequence of potentials $V_1 ,V_2, \dots V_n$, where $V_m$ has $m$
bound states, the highest $m-1$ of which are shared with $V_{m-1}$.
{}From the appendix,
these potentials may be written in terms of a set of functions $f_m$ as
\begin{equation}
V_{m}(x)= f_m^2+f_m'-\kappa_m^2
\end{equation}
or
\begin{equation}
V_{m}(x)= f_{m+1}^2-f_{m+1}'-\kappa_{m+1}^2.
\end{equation}
Using this, $V_n$ can be expressed as
\begin{equation}
V_{n}=\sum_{m=1}^{n} (V_{m}-V_{m-1})=2 {d\over{dx}} \sum_{m=1}^{n} f_m,
\end{equation}
where $V_{0}\equiv 0$.  Introducing functions $w_m$ defined
by
\begin{equation}
f_m=-w_m'/w_m,
\end{equation}
the potential $V_n$ becomes
\begin{equation}
V_n = -2  {d^2\over{dx^2}} \log \prod_{m=1}^{n} w_m,
\end{equation}
so that the $\tau$-function corresponding to $V_n$ is given by
\begin{equation}
\label{tau}
\tau_n= \prod_{m=1}^{n} w_m.
\end{equation}
We can use the operators $A_{m}=-d/dx + f_m$ to construct the functions $w_m$
from simple linear combinations of exponentials.  First observe that
the sequence of Hamiltonians $H_m = -d^2/dx^2 + V_m$ may be written as
\begin{equation}
\label{ham}
H_m= A_m^{\dagger}A_m - \kappa_m^2 = A_{m+1}A_{m+1}^{\dagger} - \kappa_{m+1}^2.
\end{equation}
Furthermore, the functions $w_m$ obey
\begin{equation}
 [A_{m}A_{m}^{\dagger}-\kappa_m^2] w_m = -\kappa_m^2 w_m.
\end{equation}
Now suppose $w_m^0$ is chosen to satisfy
\begin{equation}
\label{weqn}
- {d^2\over{dx^2}} w_m^0 = [A_{1}A_{1}^{\dagger}-\kappa_1^2]w_m^0=
-\kappa_m^2 w_m^0.
\end{equation}
Multiplying on the left by $A_{1}^{\dagger}$ and using Eq.~(\ref{ham}), we
find that $A_{1}^{\dagger} w_m^0$ obeys
\begin{equation}
 [A_{2}A_{2}^{\dagger}-\kappa_2^2]A_{1}^{\dagger}w_m^0=
-\kappa_m^2 A_{1}^{\dagger}w_m^0.
\end{equation}
Repeating this process we find that the function $w_m$ is given by
\begin{equation}
w_m = \prod_{k=1}^{n-1} A_{k}^{\dagger} w_m^0,
\end{equation}
so that the $\tau$-function can be written as
\begin{equation}
\tau_n = \prod_{m=1}^{n} \biggl{(}\prod_{k=1}^{m-1}A_{k}^{\dagger}\biggr{)}
w_m^0.
\end{equation}
Since the functions $w_m^0$ are solutions of Eq.~(\ref{weqn}), they have
the form
\begin{equation}
w_m^0= a_m^+ e^{\kappa_m x} + a_m^{-} e^{-\kappa_m x},
\end{equation}
with the constants $a_m^{\pm}$ chosen so as to ensure that $w_m$ has no
nodes.  For the case of potentials symmetric about the origin, we choose
$a_m^+ = 1$, $a_m^-=(-1)^{m+1}$.  The first few $\tau$-functions are
\begin{equation}
\label{tau0}
\tau_{0}(x)=1,
\end{equation}
\begin{equation}
\label{tau1}
\tau_{1}(x)= a_1^+ e^{\kappa_1 x} + a_1^- e^{-\kappa_1 x},
\end{equation}
\begin{eqnarray}
\label{tau2}
\tau_{2}(x)&=&a_1^+ a_2^+ (\kappa_2-\kappa_1) e^{\kappa_1 x} e^{\kappa_2 x} +
	    a_1^+ a_2^- (-\kappa_2-\kappa_1)  e^{\kappa_1 x} e^{-\kappa_2 x}
		\nonumber \\
           &+& a_1^- a_2^+ (\kappa_2+\kappa_1)  e^{-\kappa_1 x} e^{\kappa_2 x}
+
            a_1^- a_2^- (-\kappa_2+\kappa_1)  e^{-\kappa_1 x} e^{-\kappa_2 x},
\end{eqnarray}
and
\begin{eqnarray}
\label{tau3}
\tau_{3}(x)&=&a_1^+ a_2^+ a_3^ +(\kappa_3-\kappa_2)(\kappa_3-\kappa_1)
	(\kappa_2-\kappa_1) e^{\kappa_1 x+\kappa_2 x+\kappa_3 x}
\nonumber \\
	&+& a_1^+ a_2^+ a_3^ -(-\kappa_3-\kappa_2)(-\kappa_3-\kappa_1)
        (\kappa_2-\kappa_1) e^{\kappa_1 x+\kappa_2 x-\kappa_3 x} +
\nonumber \\
        &+& a_1^+ a_2^- a_3^ +(\kappa_3+\kappa_2)(\kappa_3-\kappa_1)
        (-\kappa_2-\kappa_1) e^{\kappa_1 x-\kappa_2 x+\kappa_3 x} +
\nonumber \\
        &+& a_1^+ a_2^- a_3^ -(-\kappa_3+\kappa_2)(-\kappa_3-\kappa_1)
        (-\kappa_2-\kappa_1) e^{\kappa_1 x-\kappa_2 x-\kappa_3 x} +
\nonumber \\
        &+& a_1^- a_2^+ a_3^ +(\kappa_3-\kappa_2)(\kappa_3+\kappa_1)
        (\kappa_2+\kappa_1) e^{-\kappa_1 x+\kappa_2 x+\kappa_3 x} +
\nonumber \\
        &+& a_1^- a_2^+ a_3^ -(-\kappa_3-\kappa_2)(-\kappa_3+\kappa_1)
        (\kappa_2+\kappa_1) e^{-\kappa_1 x+\kappa_2 x-\kappa_3 x} +
\nonumber \\
        &+& a_1^- a_2^- a_3^ +(\kappa_3+\kappa_2)(\kappa_3+\kappa_1)
        (-\kappa_2+\kappa_1) e^{-\kappa_1 x-\kappa_2 x+\kappa_3 x} +
\nonumber \\
        &+& a_1^- a_2^- a_3^ -(-\kappa_3+\kappa_2)(-\kappa_3+\kappa_1)
        (-\kappa_2+\kappa_1) e^{-\kappa_1 x-\kappa_2 x-\kappa_3 x}.
\end{eqnarray}
The inductive generalization to the $n$-bound state $\tau$-function is
evidently (cf.~[10])
\begin{equation}
\label{susytau}
\tau_n (x)= \prod_{m=1}^{n} \biggl{(}\biggl{[}\sum_{s_m = \pm}
	a_{m}^{s_m} e^{s_m \kappa_m x}\biggr{]}
	\prod_{l=1}^{m-1} \bigl{[}s_m \kappa_m - s_l \kappa_l\bigr{]}
	\biggl{)}.
\end{equation}

An alternative method of adding a soliton to a $\tau$-function is
provided by the vertex operator, which is a linear operator that,
when applied to a $\tau$-function, adds a soliton.
In Eqs.~(\ref{tau1},\ref{tau2},\ref{tau3}) we have displayed the first few
$\tau$-functions generated by supersymmetric quantum mechanics.  The
given expressions are somewhat deceptive, however, since they over-count
the number of degrees of freedom that one has when adding a bound state
to a potential.  From the given expressions, it would seem that one
is free to independently choose all three of the constants $a_m^{+}$,
$a_m^{-}$, and $\kappa_m$. Contrary to this, one in fact has only two
degrees of freedom:  since the multiplication of a $\tau$-function by
an overall constant leaves the potential unchanged, the true independent
variables are the ratio $a_m^{+}/a_m^{-}$ and the constant $\kappa_m$.
The former of these two is in fact a measure of the ``position'' of the
soliton that one adds when adding a bound state.  With these facts in mind,
we re-write the above $\tau$-functions in an alternative form:
\begin{equation}
\tau_{1}(x) = 1 + e^{2 \kappa_1 (x-x_1)},
\end{equation}
\begin{equation}
\tau_{2}(x) = 1 + e^{2 \kappa_1 (x-x_1)} + e^{2 \kappa_2 (x-x_2)}
	      + \biggl{(} {{\kappa_2-\kappa_1}\over{\kappa_2+\kappa_1}}
		\biggr{)} ^2 e^{2 \kappa_1 (x-x_1) + 2 \kappa_2 (x-x_2)},
\end{equation}
and,
\begin{eqnarray}
\tau_{3}(x) &=&  1 + e^{2 \kappa_1 (x-x_1)} + e^{2 \kappa_2 (x-x_2)} +
	         e^{2 \kappa_3 (x-x_3)}+
		 \biggl{(} {{\kappa_2-\kappa_1}\over{\kappa_2+\kappa_1}}
                 \biggr{)} ^2 e^{2 \kappa_1 (x-x_1) + 2 \kappa_2 (x-x_2)}
		 \nonumber \\
	    &+& \biggl{(} {{\kappa_3-\kappa_1}\over{\kappa_3+\kappa_1}}
                \biggr{)} ^2 e^{2 \kappa_1 (x-x_1) + 2 \kappa_3 (x-x_3)}+
		\biggl{(} {{\kappa_3-\kappa_2}\over{\kappa_3+\kappa_2}}
                \biggr{)} ^2 e^{2 \kappa_2 (x-x_2) + 2 \kappa_3 (x-x_3)}
		\nonumber\\
	    &+& \biggl{(} {{\kappa_2-\kappa_1}\over{\kappa_2+\kappa_1}}
                \biggr{)} ^2
		\biggl{(} {{\kappa_3-\kappa_1}\over{\kappa_3+\kappa_1}}
                \biggr{)} ^2
		\biggl{(} {{\kappa_3-\kappa_2}\over{\kappa_3+\kappa_2}}
                \biggr{)} ^2  e^{2 \kappa_1 (x-x_1) + 2 \kappa_2 (x-x_2)
				 +2 \kappa_3 (x-x_3)}.
\end{eqnarray}
In deriving these expressions from those given above, we have made use of the
fact that multiplication of a $\tau$-function by an overall constant
or by an exponential whose argument is linear in $x$ leaves the potential
unchanged.
Using these expressions, we can construct a linear operator that converts
an $n$-soliton $\tau$-function into an $n+1$-soliton $\tau$-function.
We denote this operator by $A(\kappa,\bar{x})$ to make its dependence on the
relevant variables explicit, and construct it so as to satisfy
\begin{equation}
A(\kappa,\bar{x})\tau_n = \tau_{n+1}.
\end{equation}
Certain properties of $A(\kappa,\bar{x})$ can be deduced
by considering its action on
$\tau_{0}\equiv 1$:  we require that
\begin{equation}
A(\kappa_1,x_1) \tau_{0} = A(\kappa_1,x_1) 1 = 1+e^{2\kappa_1(x-x_1)}.
\end{equation}
This equation is automatically satisfied by $A$ of the form
$A(\kappa,\bar{x})=1+B(\kappa,\bar{x})$, with $B$ constructed such that
\begin{equation}
\label{ontau0}
B(\kappa_1,x_1)1=e^{2\kappa_1(x-x_1)}.
\end{equation}
Further constraints are obtained by considering the action of
$A(\kappa,\bar{x})$ on $\tau_1$.  Using (\ref{ontau0}), and applying $A$
to $\tau_1$, we find
\begin{equation}
A(\kappa_2,x_2)\tau_1 = 1+e^{2\kappa_1(x-x_1)}+e^{2\kappa_2(x-x_2)}+
			B(\kappa_2,x_2)e^{2\kappa_1(x-x_1)}.
\end{equation}
$B$ must therefore obey
\begin{equation}
\label{bogus}
B(\kappa_2,x_2)e^{2\kappa_1(x-x_1)}=
		\biggl{(} {{\kappa_2-\kappa_1}\over{\kappa_2+\kappa_1}}
                \biggr{)} ^2 e^{2 \kappa_1 (x-x_1) + 2 \kappa_2 (x-x_2)}.
\end{equation}
Here we encounter the difficulty that such a constraint on $B$ leads to
inconsistencies when we apply it to $\tau$-functions of higher order.
A way around this difficulty [11] is suggested by
re-expressing the $\kappa$-dependent coefficient in $\tau_2$ in the
following form:
\begin{equation}
\label{expansion}
\biggl{(} {{\kappa_2-\kappa_1}\over{\kappa_2+\kappa_1}}
                \biggr{)} ^2=e^{2\bigl{(}\log(1-\kappa_1/\kappa_2)-
				\log(1+\kappa_1/\kappa_2)\bigr{)}}
		=e^{-4\sum_{k=0}^{\infty} {1\over{2k+1}}\bigl{(}
		{\kappa_1\over{\kappa_2}}\bigr{)}^{2k+1}}.
\end{equation}
We can generate precisely such a factor through the action of $B$ by
introducing a set of auxiliary variables $t_3,t_5,t_7,\dots$ in the $\tau$-
function according to the following prescription: replace the
constant of integration $x_1$ in $\tau_1$ with the ``time-dependent''
expression
$x_1(\kappa,t_3,t_5,\dots)=
\bar{x}_1-\sum_{1}^{\infty} \kappa_1^{2k}t_{2k+1}$, with $\bar{x}_1$ constant.
These auxiliary variables can then be set to zero in the final expression
for the potential.
The $\kappa$ dependent constant can be generated by making use of the
identity [based on Eq.~(\ref{expansion})]
\begin{equation}
e^{-2({1\over{\kappa_2}} \pdrv{}{x}+{1\over{3\kappa_2^3}} \pdrv{}{t_3}
+{1\over{5\kappa_2^5}} \pdrv{}{t_5}+\dots)}
e^{2\kappa_1\bigl{(}x-x_1(\kappa_1,t_1,\dots)\bigr{)}}=
\biggl{(} {{\kappa_2-\kappa_1}\over{\kappa_2+\kappa_1}}
\biggr{)}^2
e^{2\kappa_1\bigl{(}x-x_1(\kappa_1,t_1,\dots)\bigr{)}}.
\end{equation}
This is almost what we need.  The correct form of $B$ is evidently
\begin{eqnarray}
B(\kappa,\bar{x})&=& e^{2\kappa\bigl{(}x+
	\sum_{1}^{\infty} \kappa^{2k}t_{2k+1}-\bar{x}\bigr{)}}
	e^{-2\bigl{(}{1\over{\kappa}} \pdrv{}{x}
	+{1\over{3\kappa^3}} \pdrv{}{t_3}+
	{1\over{5\kappa^5}} \pdrv{}{t_5}+\dots\bigr{)}}
	\nonumber\\
	&=&e^{2\bigl{(}\sum_{k=0}^{\infty}
	\kappa^{2k+1}t_{2k+1}-\kappa\bar{x}\bigr{)}}
	e^{-2\bigl{(}\sum_{k=0}^{\infty}{1\over{(2k+1)\kappa^{2k+1}}}
	 \pdrv{}{t_{2k+1}}\bigr{)}}.
\end{eqnarray}
The latter form of $B$ has been simplified by making the identification
$x\equiv t_1$.
We see that the form of the $\kappa$ dependent factors, or ``phase shift
functions'' appearing in $\tau_{2}$ enable us to determine the form
of the vertex operator relatively easily.  This fact has been noted
elsewhere [11], in connection with the Kadomtsev-Petviashivili
equation.
The form of the operator $A$ is then
\begin{equation}
\label{vertexop}
A(\kappa,\bar x) = 1 + e^{2\bigl{(}\sum_{k=0}^{\infty}
        \kappa^{2k+1}t_{2k+1}-\kappa\bar{x}\bigr{)}}
        e^{-2\bigl{(}\sum_{k=0}^{\infty}{1\over{(2k+1)\kappa^{2k+1}}}
         \pdrv{}{t_{2k+1}}\bigr{)}}.
\end{equation}
This form of the vertex operator has been cited elsewhere [1,12],
and does in fact convert an $n$-bound state $\tau$-function into one
possessing $n+1$ bound states.  This is most readily seen by direct
computation using the identity
\begin{equation}
B(\kappa_n,\bar{x}_n)\dots B(\kappa_1,\bar{x}_1)1
=\prod_{k=1}^{n}\biggl{\{}\biggl{[}\prod_{l=1}^{k-1}
\biggl{(} {{\kappa_k-\kappa_l}\over{\kappa_k+\kappa_l}}
\biggr{)}^2\biggr{]} e^{2\bigl{(}\sum_{n=0}^{\infty}
        \kappa_{k}^{2n+1}t_{2n+1}-\kappa_{k}\bar{x}_{k}\bigr{)}} \biggl{\}}
\end{equation}
and comparing, after suitable multiplication by an overall constant and
an overall exponential factor, with the form of the $\tau$-function
given in Eq.~(\ref{susytau}).  The $n$-soliton $\tau$-function can be
generated by repeated action of the vertex operator on $\tau_0\equiv 1$:
\begin{equation}
\tau_n = A(\kappa_n,x_n)A(\kappa_{n-1},x_{n-1})\dots A(\kappa_1,x_1)1.
\end{equation}
A particularly  compact form of the vertex operator (\ref{vertexop}) can
given if we choose $\beta=\exp(-2\kappa \bar{x})$ and make use of the
fact that $B(\kappa,\bar{x})^2=0$.  We may write $A$ as
\begin{equation}
A(\kappa,\beta)=e^{\beta B(\kappa,0)}.
\end{equation}
In this form, it is apparent that the operator $B(\kappa,0)$ is a generator
of an infinitesimal symmetry that maps $\tau$-functions onto new $\tau$-
functions.  It has been shown [12] that these symmetries are generated
by a certain class of infinite-dimensional Lie algebras.
Another frequently used form of the vertex operator is
\begin{equation}
X(\kappa)= \exp{\sum_{0}^{\infty} \biggl{(}\kappa^{2k+1} t_{2k+1}\biggr{)} }
        \exp{\sum_{0}^{\infty}\biggl{(}
         -{1\over{(2k+1) \kappa^{2k+1}}}\pdrv{}{t_{2k+1}}\biggr{)}}.
\end{equation}
This form of the vertex operator can also be used to add a soliton to a
$\tau$-function: we have
\begin{equation}
\tau_n = \biggl{(} b_n^+ X(\kappa_n) + b_n^- X(-\kappa_n)\biggr{)} \tau_{n-1}.
\end{equation}
In this case, it is the ratio $b_n^+/b_n^-$ that determines the position of
the new soliton.  The equivalence of this construction to that given
above can be verified by making use of the identity
\begin{equation}
X(\kappa_n) X(\kappa_{n-1})\dots X(\kappa_1) 1=
        \prod_{m=1}^{n}
        \exp{\sum_{0}^{\infty} \biggl{(}\kappa^{2k+1}_m t_{2k+1}\biggr{)} }
        \prod_{l=1}^{m-1} \biggl{\vert}
        {{\kappa_m-\kappa_l}\over{\kappa_m+\kappa_l}}
                          \biggr{\vert}^{1/2},
\end{equation}
to construct the $n$-soliton $\tau$-function, and comparing the result, modulo
irrelevant factors, to that given above.

In the above discussion we have treated the variables $t_{2k+1}$ as
auxiliary variables which are to be set to zero after they
have been used to construct a given potential.  Comparison of
Eq.~(\ref{onesoliton}) with the vertex operator (\ref{vertexop})
shows that the auxiliary variables $t_{2k+1}$ introduced in this section
are precisely the KdV time variables employed in Sec. IV.
Consequently, the above
construction can also be used to construct multi-soliton solutions of
the KdV hierarchy.  Defining $u(x,t_3,t_5,\dots)$ by
\begin{equation}
u(x,t_3,t_5,\dots)= 2 {\partial^2\over{\partial x^2}} \log \tau(x,t_3,\dots),
\end{equation}
with $\tau$ constructed by repeated application of the vertex operator on 1,
we see that $u$ obeys
\begin{equation}
u_{t_{2k+1}}  = \pdrv{}{x} \vdrv{}{u} H_{2k+1}.
\end{equation}
As the time variables run over all possible values,  $u$
maps out all of the isospectral deformations of the potential
$V(x)=-u(x)|_{t_3,t_5,\dots {\rm fixed}}$.

\section{CONCLUSIONS}

We have explored two methods for adding a soliton to a multi-soliton solution
of the Korteweg-de Vries equation (and related higher-order equations).  The
method more familiar from the standpoint of soliton theory [1] employs a
function (the $\tau$-function) whose logarithm, when differentiated twice,
gives the solution (up to a factor).  A vertex operator, a function of
infinitely many times, acts upon this $\tau$-function to add a soliton.

The alternative method for adding a soliton relies upon supersymmetric
quantum mechanics.  In order to add a soliton, one must solve a (first-order)
Riccati equation, thereby adding one integration constant for each soliton.
The physical interpretation of this integration constant is the position
of the new soliton relative to all the others.  This position is best
visualized at large separations of the individual solitons, which occurs
at asymptotic values of the times $t_{2k+1}$ governing the evolution
according to the Hamiltonians $H_{2k+1}$.  Under such circumstances, the
multi-soliton solution closely resembles a set of individual one-soliton
``lumps,'' each lump propagating with a speed governed by its size.

We have shown that the two methods for adding solitons are equivalent.
Nonetheless, the method of demonstration still seems somewhat roundabout.
There are many features of single-soliton solutions which suggest that they
may be useful in constructing the vertex operator (\ref{vertexop}), but
we have so far been unable to find a more direct route to this result.

\section{ACKNOWLEDGMENTS}

This work was supported in part by the United States Department of
Education through a GAANN Fellowship (A.K.G.) and by the United States
Department of Energy through Grant No. DE FG02 90ER40560.

\section{APPENDIX:  DERIVATION OF THE TRANSMISSION COEFFICIENT USING
		SUPERSYMMETRIC QUANTUM MECHANICS}

We wish to show that potentials constructed using the method of Sec. II are
reflectionless and have transmission coefficients $T(k)$ given by
Eq.~(\ref{transmission}):
\begin{equation}
\label{t2}
T(k)=\prod_{i=1}^{N} {{ik-\kappa_i}\over{ik+\kappa_i}}.
\end{equation}
These results may be proven by induction on N.

For, suppose that we have constructed a series of potentials $V_{n}$,
$n=0,~1,~2,~\dots,~N$, such that $V_{n}$ has bound states at
$E=-\kappa_{1}^2,~-\kappa_{2}^2,~\dots,~-\kappa_{n}^2$.  Using the
methods of Sec. II, we may construct $V_{n+1}$ from $V_{n}$ as follows:
We write
\begin{equation}
V^{(n+1)}_{-}=V_{n}+\kappa_{n+1}^2=f_{n+1}^2-f_{n+1}^{\prime},
\end{equation}
so that the partner potential $V_{+}^{(n+1)}$ is
\begin{equation}
V^{(n+1)}_{+}=V_{n+1}+\kappa_{n+1}^2=f_{n+1}^2+f_{n+1}^{\prime}.
\end{equation}
We therefore find that each potential $V_{n}$ has two  equivalent
representations:
\begin{equation}
V_{n}=f_{n+1}^2-f_{n+1}^{\prime}-\kappa_{n+1}^{2}=
  f_{n}^2+f_{n}^{\prime}-\kappa_{n}^{2}.
\end{equation}
The eigenfunctions of $V_{n}$ can be related to those of $V_{n+1}$ using
the operators $A$ and $A^{\dagger}$.  Comparing with Sec. II, we find that
if $\psi_{n}$ is an eigenfunction of $V_{n}$, then the corresponding
eigenfunction $\psi_{n+1}$ for the potential $V_{n+1}$ is given by
\begin{equation}
\psi_{n+1}=A_{n+1}^{\dagger}\psi_{n}=\biggl{(}\deriv{}{x}+f_{n+1}\biggr{)}
	\psi_{n}.
\end{equation}
We now make the following induction hypotheses:  we suppose for some $n$ that
(i) $f_{n}\rightarrow\mp\kappa_{n}$ as $x\rightarrow\pm\infty$, and that (ii)
the transmission coefficient for $V_n$  is given by Eq.~(\ref{t2}) with $N=n$.
We know from Sec. II that both hypotheses hold for $n=1$.
It follows that $V_{n}\rightarrow 0$ as  $x\rightarrow\pm\infty$, and that
$f_{n+1}$ obeys
\begin{equation}
f_{n+1}^2-f_{n+1}^{\prime}=V_{n}+\kappa_{n+1}^2.
\end{equation}
As $x\rightarrow\pm\infty$, this reduces to $f_{n+1}^2-f_{n+1}^{\prime}
=\kappa_{n+1}^2$,
which has the solution $f_{n+1}=-\kappa_{n+1}\tanh\kappa_{n+1}(x-x_0 )$.
Consequently $f_{n+1}$ has the asymptotic forms
\begin{equation}
f_{n+1}\rightarrow\cases{
                        -\kappa_{n+1}\tanh\kappa_{n+1}(x-x_{+})\rightarrow
			-\kappa_{n+1},  & $x \rightarrow +\infty$;\cr
\cr
			 -\kappa_{n+1}\tanh\kappa_{n+1}(x-x_{-})\rightarrow
                        \kappa_{n+1},  & $x \rightarrow -\infty$.\cr
}
\end{equation}
This proves that our first induction hypothesis holds for all $n$.  To see
that the second holds, observe that a plane wave solution in the potential
$V_{n}$ has the asymptotic form
\begin{equation}
\psi_{n}(x)\rightarrow\cases{
                        e^{ikx},  & $x \rightarrow -\infty$;\cr
\cr
                        \prod_{i=1}^{n} {{ik-\kappa_i}\over{ik+\kappa_i}}
			e^{ikx}, & $x \rightarrow +\infty$.\cr}
\end{equation}
Applying the operator $A_{n+1}^{\dagger}$ to $\psi_n$ gives the corresponding
solution for the potential $V_{n+1}$.  After dividing through by
$ik+\kappa_{n+1}$, we find that this solution has the form
\begin{equation}
\psi_{n+1}(x)\rightarrow\cases{
                        e^{ikx},  & $x \rightarrow -\infty$;\cr
\cr
                        \prod_{i=1}^{n+1} {{ik-\kappa_i}\over{ik+\kappa_i}}
                        e^{ikx}, & $x \rightarrow +\infty$,\cr}
\end{equation}
which proves the second of our induction hypotheses, as well as Eq.~(\ref{t2}).

\newpage
\centerline{\bf{REFERENCES}}
\medskip

\begin{enumerate}

\item[{[1]}] Alan C. Newell, {\it Solitons in Mathematics and Physics} (SIAM,
Philadelphia, 1985).

\item[{[2]}] E. Witten, Nucl. Phys. {\bf B188}, 513 (1981).

\item[{[3]}] C. V. Sukumar, J. Phys. A {\bf 18}, L57 (1985); {\bf 18}, 2917
(1985); {\bf 18}, 2937 (1985); {\bf 19}, 2229 (1986); {\bf 19}, 2297 (1986);
{\bf 20}, 2461 (1987).

\item[{[4]}] Waikwok Kwong and Jonathan L. Rosner, Prog. Theor. Phys. Suppl.
{\bf 86}, 366 (1986) (Festschrift volume in honor of Y. Nambu).

\item[{[5]}] Waikwok Kwong, Harold Riggs, Jonathan L. Rosner and H. B. Thacker,
Phys. Rev. D {\bf 39} 1242 (1989).

\item[{[6]}] Qin-mou Wang, Uday P. Sukhatme, Wai-Yee Keung, and Tom D. Imbo,
Mod. Phys. Lett. A {\bf 5}, 525 (1990).

\item[{[7]}] M. Adler and J. Moser, Comm. Math. Phys. {\bf 61}, 1 (1978);
B. A. Kuperschmidt and G. Wilson, Inventiones Math. {\bf 62}, 403 (1981).

\item[{[8]}] H. Flaschka, Quart. J. Math. Oxford {\bf 34}, 61 (1983).

\item[{[9]}] R. M. Miura, C. S. Gardner, and M. D. Kruskal, J. Math. Phys.
{\bf 9}, 1204 (1968).

\item[{[10]}] Jonathan F. Schonfeld, Waikwok Kwong, Jonathan L. Rosner, C.
Quigg, and H. B. Thacker, Ann. Phys. (N.Y.) {\bf 128}, 1 (1980).

\item[{[11]}] Ryogo Hirota,
in {\it Non-linear Integrable Systems -- Classical and Quantum Theory},
Kyoto, May 13--16 1981, ed. by M. Jimbo and T. Miwa, p. 17.

\item[{[12]}] Etsuro Date, Michio Jimbo,  Masaki Kashiwara, and Tetsuji Miwa
in {\it Non-linear Integrable Systems -- Classical and Quantum Theory},
Kyoto, May 13--16 1981, ed. by M. Jimbo and T. Miwa, p. 39.

\end{enumerate}

\end{document}